\documentclass[12pt,a4paper]{article}
\usepackage{amsmath}
\usepackage[font={footnotesize,it}]{caption}
\usepackage[top=1.3in, bottom=1.3in, left=1.2in, right=1.2in]{geometry}

\DeclareCaptionStyle{italic}[justification=centering]{labelfont={bf},textfont={it},labelsep=colon}
\usepackage{graphicx,psfrag,epsf}
\usepackage{enumerate}
\usepackage{natbib}
\usepackage{url, setspace}
\usepackage{booktabs, subfig, bm, paralist,mathpazo,tikz,todonotes,longtable,microtype,dsfont,rotating} 
\usepackage[pdftex,colorlinks=true]{hyperref}
\definecolor{darkblue}{rgb}{0,0,.6}
\hypersetup{citecolor=darkblue,linkcolor=darkblue,urlcolor=darkblue}

\newcommand{\blind}{0}

\graphicspath{{plots/}}

\newsavebox\CBox
\def\textBF#1{\sbox\CBox{#1}\resizebox{\wd\CBox}{\ht\CBox}{\textbf{#1}}}

\date{\today}

\begin{document}

\def\spacingset#1{\renewcommand{\baselinestretch}
{#1}\small\normalsize} \spacingset{1}

\if0\blind
{
  \title{\bf Retiree mortality forecasting: \mbox{A partial age-range or a full age-range model?}}
  \author{Han Lin Shang\thanks{Postal address: Department of Actuarial Studies and Business Analytics, Level 7, 4 Eastern Road, Macquarie University, NSW 2109, Australia; Email: hanlin.shang@mq.edu.au; ORCID: \url{https://orcid.org/0000-0003-1769-6430}}
  \hspace{.2cm}\\
    Department of Actuarial Studies and Business Analytics \\
    Macquarie University \\
    \\
    Steven Haberman \\
    Cass Business School \\
    City, University of London}
  \maketitle
} \fi

\if1\blind
{
   \title{\bf Retiree mortality forecasting: \mbox{A partial age-range or a full age-range model?}}
   \author{}
   \maketitle
} \fi

\bigskip

\begin{abstract}
An essential input of annuity pricing is the future retiree mortality. From observed age-specific mortality data, modeling and forecasting can be taken place in two routes. On the one hand, we can first truncate the available data to retiree ages and then produce mortality forecasts based on a partial age-range model. On the other hand, with all available data, we can first apply a full age-range model to produce forecasts and then truncate the mortality forecasts to retiree ages. We investigate the difference in modeling the logarithmic transformation of the central mortality rates between a partial age-range and a full age-range model, using data from mainly developed countries in the \cite{HMD19}. By evaluating and comparing the short-term point and interval forecast accuracies, we recommend the first strategy by truncating all available data to retiree ages and then produce mortality forecasts. However, when considering the long-term forecasts, it is unclear which strategy is better since it is more difficult to find a model and parameters that are optimal. This is a disadvantage of using methods based on time series extrapolation for long-term forecasting. Instead, an expectation approach, in which experts set a future target, could be considered, noting that this method has also had limited success in the past.

\vspace{.2in}
\end{abstract}

\newpage
\spacingset{1.46}

\section{Introduction}

Improving human survival probability contributes greatly to an aging population. To guarantee one individual's financial income in retirement, a policyholder may purchase a fixed-term or lifetime annuity. A fixed-term or lifetime annuity is a contract offered by insurers guaranteeing regular payments in exchange for an initial premium. Since an annuity depends on survival probabilities and interest rates, pension funds and insurance companies are more likely to face a risk of longevity. Longevity risk is a potential systematic risk attached to the increasing life expectancy of policyholders, which can eventually result in a higher payout ratio than expected \citep{CDR08}. The concerns about longevity risk have led to a surge of interest in modeling and forecasting age-specific mortality rates. 

Many models for forecasting age-specific mortality indicators have been proposed in demographic literature \citep[see][for reviews]{BT08}. Of these, \cite{LC92} implemented a principal component method to model the logarithm of age-specific mortality rates ($m_{x,t}$) and extracted a single time-varying index representing the trend in the level of mortality, from which the forecasts are obtained by a random walk with drift. Since then, the Lee-Carter (LC) method has been extended and modified \citep[see][for reviews]{BT08, PDH+09, SBH11, SH18}. The LC method has been applied to many countries, including Belgium \citep{BDV02}, Austria \citep{CP01}, England and Wales \citep{RH03}, and Spain \citep{GV05, DMP08}.

Many statistical models focus on time-series extrapolation of past trends exhibited in age-specific mortality rates \citep[see, e.g.,][]{BT08}. We consider two modeling strategies: On the one hand, we can first truncate all available data to retiree ages and then produce mortality forecasts \citep[see, e.g.,][]{CBD06, CBD+09}. On the other hand, we can first use the available data to produce forecasts and then truncate the mortality forecasts to retiree ages \citep[see, e.g.,][]{SH17}. In this paper, our contribution is to investigate the difference in modeling the logarithmic transformation of the central mortality rates between a partial age-range and a full age-range model, using data from mainly developed countries in the \cite{HMD19}. By evaluating and comparing the short-term point and interval forecast accuracies, we recommend the first strategy by truncating all available data to retiree ages and then produce mortality forecasts. However, when we consider the long-term forecasts, it is unclear which strategy is better since it is more difficult to find a model and parameters that are optimal. This is a disadvantage of using methods based on time series extrapolation for long-term forecasting. Instead, an expectation approach, in which experts set a future target, could be considered, noting that this method has also had limited success in the past \citep{BT08}. Our recommendations could be useful to actuaries for choosing a better modeling strategy and more accurately pricing a range of annuity products.

The article is organized as follows: In Section~\ref{sec:2}, we describe the mortality data sets of 19 mainly developed countries. We revisit ﬁve time-series extrapolation models for forecasting age-speciﬁc mortality rates, which have been shown in the literature to work well across the full age range for some data sets \citep[for more details, consult][]{Shang12, SH18}. Using these models as a testbed, we compare point and interval forecast accuracies between the two modeling strategies and provide our recommendations in Section~\ref{sec:5}. Conclusions are presented in Section~\ref{sec:6}.

\section{Data sets}\label{sec:2}

The data sets used in this study were taken from the \cite{HMD19}. For each sex in a given calendar year, the mortality rates obtained by the ratio between ``number of deaths" and ``exposure to risk" are arranged in a matrix for age and calendar year. Nineteen countries, mainly developed countries, were selected, and thus 38 sub-populations of age- and sex-specific mortality rates were obtained for all analyses. The 19 countries selected all have reliable data series commencing at/before 1950. Due to possible structural breaks (i.e., two world wars), we truncate all data series from 1950 onwards. The omission of Germany is because the \cite{HMD19} for a reunited Germany only dates back to 1990. The selected countries and their abbreviations are shown in Table~\ref{fig:1}, along with their last year of available data (recorded in April 2019). To avoid fluctuations at older ages, we consider ages from 0 to 99 in a single year of age and the last age group is from 100 onwards. Should we consider all ages from 0 to 110+, and we may encounter the missing-value issue and observe mortality rates outside the range of $[0,1]$ for some years.

\begin{table}[!htbp]
\centering
\caption{The 19 countries with the initial year of 1950 and their ending year listed below.}\label{fig:1}
\tabcolsep 0.07in
\begin{tabular}{@{}llllll@{}}
\toprule
Country & Abbreviation & Last year  & Country & Abbreviation & Last year \\
\midrule
Australia & AUS & 2014 & Norway & NOR & 2014 \\ 
Belgium & BEL & 2015 & Portgual & PRT & 2015 \\
Canada & CAN & 2011 & Spain & SPA & 2016 \\
Denmark & DEN & 2016 & Sweden & SWE & 2016 \\
Finland & FIN & 2015 & Switzerland & SWI & 2016 \\
France & FRA & 2016 & Scotland & SCO &  2016 \\
Italy & ITA & 2014 & England \& Wales & EW & 2016 \\
Japan & JPN & 2016 & Ireland & IRE & 2014\\
Netherland & NET & 2016 & United States of America & USA & 2016\\
New Zealand & NZ & 2013 \\  
\bottomrule
\end{tabular}
\end{table}

\section{Results}\label{sec:5}

\subsection{Forecast evaluation}

We present 19 countries that begin in 1950 and end in the last year listed in Table~\ref{fig:1}. We keep the last 30 observations for forecasting evaluation, while the remaining observations are treated as initial fitting observations, from which we produce the one-step-ahead to 30-step-ahead forecasts. Via an expanding window approach \citep[see also][]{ZW06}, we re-estimate the parameter in the time series forecasting models by increasing the fitted observations by one year and produce the one-step-ahead to 29-step-ahead forecasts. We iterate this process by increasing the sample size by one year until the end of the data period. The process produces 30 one-step-ahead forecasts, 29 two-step-ahead forecasts, $\dots$, one 30-step-ahead forecast. We compare these forecasts with the holdout samples to determine the out-of-sample forecast accuracy.

\subsection{Forecast error criteria}

To evaluate the point forecast accuracy, we consider the mean absolute percentage error (MAPE) and root mean squared percentage error (RMSPE). The MAPE and RMSPE criteria measure how close the forecasts compare with the actual values of the variable being forecast, regardless of the error sign. The MAPE and RMSPE criteria can be expressed as:
\begin{align*}
\text{MAPE}_h &= \frac{1}{p\times (31-h)}\sum^{(31-h)}_{j=1}\sum_{x=1}^{p}\left|\frac{m_{x,j} - \widehat{m}_{x,j}}{m_{x,j}}\right|\times 100,\qquad x=1,\dots,p, \\
\text{RMSPE}_h &= \sqrt{\frac{1}{p\times (31-h)}\sum^{(31-h)}_{j=1}\sum^p_{x=1}\left(\frac{m_{x,j} - \widehat{m}_{x,j}}{m_{x,j}}\right)^2\times 100}
\end{align*}
where $m_{x,j}$ represents the actual holdout sample for age $x$ in the forecasting year $j$, $p$ denotes the total number of ages, and $\widehat{m}_{x,j}$ represents the forecasts for the holdout sample.

To evaluate the pointwise interval forecast accuracy, we consider the interval score criterion of \cite{GR07}. We consider the common case of the symmetric $100(1-\alpha)\%$ prediction intervals, with lower and upper bounds that were predictive quantiles at $\alpha/2$ and $1-\alpha/2$, denoted by $\widehat{m}_{x, j}^{\text{lb}}$ and $\widehat{m}_{x, j}^{\text{ub}}$. As defined by \cite{GR07}, a scoring rule for evaluating the pointwise interval forecast accuracy at time point $j$ is
\begin{align*}
S_{\alpha}\left(\widehat{m}_{x,j}^{\text{lb}}, \widehat{m}_{x, j}^{\text{ub}}; m_{x, j}\right) = \left(\widehat{m}_{x, j}^{\text{ub}} - \widehat{m}_{x, j}^{\text{lb}}\right) + & \frac{2}{\alpha}\left(\widehat{m}_{x,j}^{\text{lb}} - m_{x, j}\right)\mathds{1}\left\{m_{x,j} < \widehat{m}_{x, j}^{\text{lb}}\right\} \\
 &\frac{2}{\alpha}\left(m_{x, j} - \widehat{m}_{x, j}^{\text{ub}}\right)\mathds{1}\left\{m_{x, j} > \widehat{m}_{x, j}^{\text{ub}}\right\},
\end{align*}
where $\mathds{1}\{\cdot\}$ denotes the binary indicator function. The optimal interval score is achieved when $m_{x, j}$ lies between $\widehat{m}_{x,j}^{\text{lb}}$ and $\widehat{m}_{x, j}^{\text{ub}}$, with the distance between the upper and lower bounds being minimal. To obtain summary statistics of the interval score, we take the mean interval score across different ages and forecasting years. The mean interval score can be expressed as:
\begin{equation*}
\overline{S}_{\alpha, h} = \frac{1}{p\times (31-h)}\sum^{(31-h)}_{j=1}\sum^p_{x=1}S_{\alpha}\left(\widehat{m}_{x, j}^{\text{lb}}, \widehat{m}_{x, j}^{\text{ub}}; m_{x, j}\right).
\end{equation*}

\subsection{Comparison of point and interval forecast errors}

Modeling and forecasting mortality can be taken place in two routes. On the one hand, we can first truncate the available data to certain ages, such as from 60 to 99 in a single year of age and 100+ as the last age, and then produce mortality forecasts for these retiree ages. On the other hand, we can first use the available data, i.e., age-specific mortality from 0 to 99 in a single year of age and 100+ as the last age, to produce forecasts for these 101 ages and then truncate the mortality forecasts to certain ages, such as 60 to 99 in a single year of age and 100+ as the last age. 

We study five time-series extrapolation models for forecasting age-specific mortality, which have been shown in the literature to work well across the full age range for some data sets. Note that the Cairns-Blake-Dowd suite of models are not included in this paper, because they are designed just for ages 55 and over. The models that we have considered are subjective and far from extensive, but they suffice to serve as a testbed for comparing the forecast accuracy. These five models are: the Lee-Carter model with Poisson errors \citep[see, e.g.,][]{BDV02, RH03, RH06}, the Lee-Carter model with Gaussian errors \citep[see, e.g.,][]{BMS02, RH03, KSH06}, age-period-cohort model \citep[see, e.g.,][]{RH06} and the Plat model \citep{Plat09}. 

For the short-term forecast horizon (i.e., the one-step-ahead forecast horizon), we compute the mean of the MAPEs and RMSPEs to evaluate the point forecast accuracy. From Table~\ref{tab:1}, there is an advantage of directly modeling and forecasting the truncated series for the female mortality. For modeling the male mortality, the advantage of directly modeling and forecasting the truncated series intensifies. By comparing the mean errors of the 19 countries, Table~\ref{tab:1} shows that the most accurate forecasting method is the Plat model for providing best estimates of the female and male mortality forecasts. The Lee-Carter model with Poisson errors produces smaller MAPEs and RMSPEs than the Lee-Carter model with Gaussian errors. With the Lee-Carter model with Gaussian errors, it is advantageous to consider two components rather than only the first component.

\begin{center}
\tabcolsep 0.06in
\begin{longtable}{@{}llrrrrrrrrrrrr@{}}
\caption{For the one-step-ahead forecast horizon $h=1$, we compute the MAPE and RMSPE for each country and each model. The most accurate model for each country is highlighted in bold. For each model, we consider modeling the data with either a partial age range (termed as Partial) or a full age range (termed as Full).} \label{tab:1} \\
\toprule
Error & Country & \multicolumn{2}{c}{LC (Poisson)} & \multicolumn{2}{c}{LC (Gaussian)}  & \multicolumn{2}{c}{LC$_2$ (Gaussian)} & \multicolumn{2}{c}{APC} & \multicolumn{2}{c}{Plat} \\ 
& & Full & Partial  & Full & Partial   & Full & Partial   & Full & Partial   & Full & Partial   \\
\midrule
\endfirsthead

\toprule
Error & Country & \multicolumn{2}{c}{LC (Poisson)} & \multicolumn{2}{c}{LC (Gaussian)}  & \multicolumn{2}{c}{LC$_2$ (Gaussian)} & \multicolumn{2}{c}{APC} & \multicolumn{2}{c}{Plat} \\ 
& & Full & Partial  & Full & Partial   & Full & Partial   & Full & Partial   & Full & Partial \\
 \midrule
\endhead

\hline \multicolumn{12}{r}{{Continued on next page}} \\ 
\endfoot
\endlastfoot
MAPE & \underline{Female} & \\
& AUS & 4.45 & \textBF{4.38} & 6.01 & 4.97 & 4.69 & 4.42 & 7.32 & 6.38 & 4.83 & 4.66 \\ 
&  BEL & 4.92 & 4.90 & 6.08 & 5.01 & 5.84 & 5.04 & 7.85 & 6.24 & \textBF{4.44} & 4.74 \\ 
&  CAN & 3.29 & 3.22 & 3.94 & 3.57 & 3.74 & 3.03 & 4.98 & 4.29 & 3.20 & \textBF{2.73} \\ 
&  DEN & 7.05 & 7.15 & 8.25 & 6.90 & 7.74 & 6.36 & 7.58 & 7.15 & 5.31 & \textBF{4.79} \\ 
&  FIN & 6.43 & 6.13 & 11.56 & 6.80 & 8.26 & 6.80 & 9.31 & 7.94 & \textBF{5.63} & 5.73 \\ 
&  FRA & 4.28 & 4.47 & 4.97 & 4.47 & 3.96 & 3.80 & 7.67 & 6.16 & 4.22 & \textBF{3.65} \\ 
&  ITA & 3.25 & 3.15 & 8.36 & 3.71 & \textBF{3.12} & 3.17 & 6.49 & 5.56 & 3.68 & 3.43 \\ 
&  JPN & 5.99 & 6.02 & 23.08 & 7.03 & 3.84 & \textBF{3.23} & 6.00 & 5.02 & 4.04 & 3.31 \\ 
&  NET & 3.62 & \textBF{3.61} & 4.58 & 3.67 & 4.39 & 3.65 & 7.53 & 6.29 & 4.19 & 3.78 \\ 
&  NZ & \textBF{7.70} & 7.75 & 8.65 & 7.90 & 8.33 & 7.85 & 10.02 & 9.38 & 7.75 & 7.74 \\ 
&  NOR & \textBF{5.25} & 5.28 & 6.92 & 5.54 & 6.41 & 5.41 & 8.18 & 7.58 & 5.66 & 5.56 \\ 
&  PRT & 6.06 & \textBF{5.42} & 11.68 & 5.84 & 7.82 & 5.78 & 8.75 & 7.30 & 6.13 & 6.58 \\ 
&  SPA & 5.15 & 5.04 & 12.99 & 5.73 & 6.03 & \textBF{4.19} & 7.92 & 6.24 & 4.80 & 5.21 \\ 
&  SWE & \textBF{4.03} & \textBF{4.03} & 4.31 & 4.09 & 4.55 & 4.11 & 8.19 & 7.00 & 4.91 & 4.31 \\ 
&  SWI & 4.69 & 4.72 & 5.99 & \textBF{4.65} & 5.89 & 4.84 & 8.94 & 7.39 & 5.28 & 5.03 \\ 
&  SCO & 6.79 & 6.83 & 8.78 & 6.55 & 7.62 & 6.49 & 7.21 & 6.97 & 5.48 & \textBF{5.15} \\ 
&  EW & 4.79 & 4.72 & 6.27 & 4.49 & 5.20 & 3.51 & 6.32 & 5.80 & 3.92 & \textBF{3.15} \\ 
&  IRE & 8.45 & \textBF{7.33} & 15.01 & 8.27 & 9.87 & 8.01 & 9.52 & 9.52 & 9.02 & 9.19 \\ 
&  USA & 3.33 & 3.22 & 3.84 & 3.20 & 3.81 & 2.37 & 4.50 & 4.28 & 3.43 & \textBF{2.13} \\ 
 \cmidrule{2-12}
&  Mean & 5.24 & 5.12 & 8.49 & 5.39 & 5.85 & 4.85 & 7.59 & 6.66 & 5.05 & \textBF{4.78} \\ 
\\
&  \underline{Male} & \\
& AUS & 5.39 & \textBF{4.84} & 7.85 & 5.39 & 5.74 & 5.58 & 7.71 & 5.98 & 5.15 & 5.27 \\ 
&   BEL & 7.86 & 6.11 & 10.53 & 9.74 & 8.43 & 7.43 & 6.29 & 5.44 & 5.31 & \textBF{4.86} \\ 
&   CAN & 5.54 & 4.28 & 7.58 & 5.05 & 5.85 & 4.76 & 5.19 & 4.25 & 4.12 & \textBF{3.31} \\ 
&   DEN & 8.19 & 6.67 & 10.29 & 10.51 & 7.77 & 8.33 & 7.77 & 6.53 & 5.78 & \textBF{5.52} \\ 
&   FIN & 8.12 & 7.49 & 11.84 & 8.57 & 10.02 & 8.43 & 8.11 & 7.10 & 6.61 & \textBF{6.59} \\ 
&   FRA & 3.71 & \textBF{3.17} & 4.42 & 3.96 & 4.11 & 3.66 & 7.54 & 6.02 & 4.16 & 3.32 \\ 
&   ITA & 5.85 & 4.21 & 11.29 & 5.96 & 4.63 & 3.73 & 5.16 & 4.96 & 3.74 & \textBF{2.87} \\ 
&   JPN & 3.95 & 3.76 & 9.58 & 4.31 & 3.78 & 4.13 & 4.92 & 4.61 & 3.59 & \textBF{3.27} \\ 
&   NET & 8.02 & 5.58 & 12.65 & 7.75 & 7.23 & 5.37 & 5.83 & 4.79 & 5.24 & \textBF{4.03} \\ 
&   NZ & 9.21 & 8.17 & 12.63 & 10.86 & 9.87 & 10.90 & 9.78 & 8.79 & \textBF{8.12} & 8.38 \\ 
&   NOR & 8.66 & 6.37 & 13.61 & 6.84 & 7.21 & 6.85 & 7.42 & 6.59 & 6.07 & \textBF{6.02} \\ 
&   PRT & 6.12 & \textBF{5.57} & 8.70 & 7.26 & 8.46 & 6.30 & 7.44 & 6.90 & 5.85 & 5.88 \\ 
&   SPA & 4.39 & \textBF{3.88} & 8.48 & 5.44 & 7.10 & 4.38 & 6.52 & 5.80 & 4.11 & 4.29 \\ 
&   SWE & 5.05 & 4.51 & 9.82 & 5.38 & 6.01 & 5.28 & 8.07 & 6.02 & 4.53 & \textBF{4.08} \\ 
&   SWI & 5.62 & 5.37 & 6.51 & 5.76 & 6.86 & 5.99 & 7.71 & 6.55 & \textBF{5.27} & 5.30 \\ 
&   SCO & 7.06 & 6.20 & 10.63 & 8.56 & 8.42 & 7.51 & 7.51 & 6.71 & 6.29 & \textBF{5.80} \\ 
&   EW & 4.78 & 4.04 & 8.52 & 5.15 & 5.99 & 4.28 & 6.73 & 4.93 & 3.49 & \textBF{3.43} \\ 
&   IRE & 13.24 & \textBF{9.34} & 17.28 & 12.43 & 12.59 & 10.55 & 10.07 & 9.84 & 9.86 & 9.71 \\ 
&   USA & 4.65 & 3.75 & 5.90 & 4.33 & 5.62 & 3.31 & 5.79 & 4.26 & 3.71 & \textBF{2.42} \\ 
\cmidrule{2-12}
&   Mean & 6.60 & 5.44 & 9.90 & 7.01 & 7.14 & 6.14 & 7.14 & 6.11 & 5.26 & \textBF{4.97} \\ 
\cmidrule{2-12}
RMSPE & \underline{Female} & \\
& AUS & 5.84 & \textBF{5.77} & 7.62 & 6.49 & 6.11 & 5.82 & 8.86 & 7.87 & 6.20 & 6.19 \\ 
&  BEL & 6.40 & 6.43 & 7.55 & 6.39 & 7.37 & 6.47 & 9.46 & 7.64 & \textBF{5.74} & 6.30 \\ 
&  CAN & 4.17 & 4.11 & 4.96 & 4.59 & 4.74 & 3.98 & 6.06 & 5.31 & 3.96 & \textBF{3.50} \\ 
&  DEN & 9.36 & 9.70 & 10.05 & 9.23 & 9.55 & 8.41 & 9.32 & 8.83 & 6.67 & \textBF{6.10} \\ 
&  FIN & 8.51 & 8.32 & 14.54 & 8.85 & 10.88 & 9.14 & 11.69 & 10.12 & \textBF{7.46} & 8.05 \\ 
&  FRA & 5.57 & 5.87 & 6.22 & 5.51 & 5.03 & \textBF{4.85} & 9.15 & 7.53 & 5.33 & 4.94 \\ 
&  ITA & 4.14 & \textBF{4.01} & 9.80 & 4.62 & 4.05 & 4.19 & 7.90 & 6.92 & 4.75 & 4.59 \\ 
&  JPN & 7.13 & 7.54 & 25.03 & 8.26 & 4.78 & \textBF{4.07} & 7.32 & 6.30 & 5.00 & 4.51 \\ 
&  NET & 4.62 & \textBF{4.60} & 5.74 & 4.73 & 5.50 & 4.68 & 8.86 & 7.55 & 5.23 & 4.88 \\ 
&  NZ & 10.41 & 10.44 & 11.48 & 10.55 & 11.16 & 10.67 & 12.75 & 12.13 & \textBF{10.29} & 10.49 \\ 
&  NOR & \textBF{6.85} & 6.92 & 8.63 & 7.20 & 8.13 & 7.13 & 10.15 & 9.43 & 7.23 & 7.35 \\ 
&  PRT & 7.85 & \textBF{7.03} & 13.55 & 7.47 & 9.85 & 7.51 & 10.93 & 9.16 & 8.28 & 8.71 \\ 
&  SPA & 6.53 & 6.51 & 14.70 & 7.18 & 7.85 & \textBF{5.26} & 9.59 & 7.60 & 6.22 & 6.86 \\ 
&  SWE & 5.27 & \textBF{5.26} & 5.54 & 5.31 & 5.95 & 5.38 & 9.75 & 8.48 & 6.13 & 5.56 \\ 
&  SWI & 5.99 & 6.03 & 7.55 & \textBF{5.91} & 7.48 & 6.12 & 10.63 & 8.98 & 6.63 & 6.54 \\ 
&  SCO & 8.96 & 8.99 & 11.57 & 8.58 & 9.86 & 8.56 & 9.07 & 8.89 & 7.11 & \textBF{6.77} \\ 
&  EW & 6.33 & 6.25 & 8.29 & 5.94 & 6.74 & 4.42 & 7.69 & 7.15 & 4.86 & \textBF{3.96} \\ 
&  IRE & 11.10 & \textBF{9.63} & 18.75 & 10.82 & 12.88 & 10.58 & 12.22 & 12.43 & 11.76 & 12.26 \\ 
&  USA & 4.19 & 4.13 & 4.79 & 4.13 & 4.73 & 3.27 & 5.91 & 5.65 & 4.29 & \textBF{2.77} \\ 
\cmidrule{2-12}
&  Mean & 6.80 & 6.71 & 10.34 & 6.94 & 7.51 & 6.34 & 9.33 & 8.31 & 6.48 & \textBF{6.33} \\ 
\\
&  \underline{Male} & \\
& AUS & 7.21 & \textBF{6.78} & 9.72 & 7.29 & 7.42 & 7.24 & 9.67 & 7.83 & 7.12 & 7.48 \\ 
&   BEL & 9.98 & 8.47 & 13.14 & 13.50 & 10.48 & 10.02 & 8.86 & 7.83 & 7.39 & \textBF{7.19} \\ 
&   CAN & 6.80 & 5.48 & 9.35 & 6.22 & 7.37 & 6.02 & 6.77 & 5.67 & 5.19 & \textBF{4.42} \\ 
&   DEN & 11.05 & 9.23 & 13.50 & 14.33 & 10.49 & 11.59 & 10.20 & 8.75 & 7.97 & \textBF{7.86} \\ 
&   FIN & 13.16 & 13.07 & 15.27 & 12.49 & 15.45 & 13.39 & 11.84 & 10.88 & \textBF{10.56} & 11.04 \\ 
&   FRA & 4.72 & \textBF{4.11} & 5.60 & 5.01 & 5.12 & 4.75 & 9.21 & 7.68 & 5.41 & 4.45 \\ 
&   ITA & 7.17 & 5.35 & 13.72 & 7.28 & 6.31 & 4.84 & 6.83 & 6.50 & 4.88 & \textBF{4.04} \\ 
&   JPN & 4.91 & 4.78 & 10.98 & 5.45 & 4.77 & 5.17 & 6.20 & 5.93 & 4.45 & \textBF{4.38} \\ 
&   NET & 9.68 & 7.05 & 15.47 & 9.82 & 8.96 & 6.80 & 7.61 & 6.36 & 5.60 & \textBF{5.54} \\ 
&   NZ & 13.36 & 12.08 & 16.86 & 15.04 & 13.59 & 14.72 & 12.94 & 11.90 & \textBF{11.72} & 12.27 \\ 
&   NOR & 11.32 & 8.98 & 17.02 & 9.45 & 9.82 & 9.73 & 9.89 & 8.98 & \textBF{8.42} & 8.62 \\ 
&   PRT & 8.11 & \textBF{7.59} & 10.55 & 9.59 & 10.56 & 8.60 & 9.72 & 9.07 & 8.06 & 8.46 \\ 
&   SPA & 5.70 & \textBF{5.31} & 9.97 & 7.01 & 9.17 & 5.73 & 8.23 & 7.71 & 5.47 & 5.64 \\ 
&   SWE & 6.56 & 6.04 & 11.71 & 6.99 & 7.75 & 7.04 & 9.72 & 7.66 & 5.89 & \textBF{5.60} \\ 
&   SWI & 7.56 & 7.30 & 8.50 & 7.77 & 9.24 & 8.19 & 10.02 & 8.60 & \textBF{7.16} & 7.34 \\ 
&   SCO & 9.21 & 8.43 & 12.98 & 10.73 & 11.10 & 10.28 & 10.04 & 9.05 & 8.50 & \textBF{7.97} \\ 
&   EW & 5.93 & 5.16 & 10.38 & 6.34 & 7.42 & 5.41 & 8.39 & 6.37 & \textBF{4.50} & 4.59 \\ 
&   IRE & 19.05 & \textBF{14.78} & 23.17 & 17.45 & 18.24 & 15.44 & 15.49 & 14.90 & 15.19 & 15.23 \\ 
&   USA & 5.59 & 4.82 & 6.95 & 5.30 & 6.66 & 4.08 & 7.87 & 6.14 & 4.78 & \textBF{3.26} \\ 
\cmidrule{2-12}
& Mean & 8.79 & 7.62 & 12.36 & 9.32 & 9.47 & 8.37 & 9.45 & 8.31 & 7.28 & \textBF{7.13} \\ 
\bottomrule
\end{longtable}
\end{center}

For the short-term forecast horizon (i.e., the one-step-ahead forecast horizon), we compute the $\overline{S}_{\alpha=0.2}$ to evaluate the interval forecast accuracy. From Table~\ref{tab:3}, there is an advantage of directly modeling and forecasting the truncated series for both female and male mortality. By comparing the mean errors of the 19 countries, Table~\ref{tab:3} shows that the most accurate forecasting method is the Plat model for providing the interval forecasts of both female and male mortality rates.

\begin{center}
\begin{longtable}{@{}lrrrrrrrrrrrr@{}}
\caption{For the one-step-ahead forecast horizon $h=1$, we compute the mean interval score ($\times 100$) for each country and each model.} \label{tab:3} \\
\toprule
Country & \multicolumn{2}{c}{LC (Poisson)} & \multicolumn{2}{c}{LC (Gaussian)}  & \multicolumn{2}{c}{LC$_2$ (Gaussian)} & \multicolumn{2}{c}{APC} & \multicolumn{2}{c}{Plat} \\ 
 & Full & Partial  & Full & Partial   & Full & Partial   & Full & Partial   & Full & Partial \\
 \midrule
\endfirsthead

\toprule
Country & \multicolumn{2}{c}{LC (Poisson)} & \multicolumn{2}{c}{LC (Gaussian)}  & \multicolumn{2}{c}{LC$_2$ (Gaussian)} & \multicolumn{2}{c}{APC} & \multicolumn{2}{c}{Plat} \\ 
 & Full & Partial  & Full & Partial   & Full & Partial   & Full & Partial   & Full & Partial \\
  \midrule
\endhead

\hline \multicolumn{11}{r}{{Continued on next page}} \\ 
\endfoot
\endlastfoot
\underline{Female} & \\
AUS & 3.51 & 3.44 & 2.70 & 2.68 & 2.59 & 2.78 & 6.60 & 5.26 & 2.70 & \textBF{2.53} \\ 
  BEL & 3.58 & 3.28 & 3.34 & 3.11 & 3.28 & 3.19 & 6.09 & 4.68 & \textBF{2.61} & 2.64 \\ 
  CAN & 2.91 & 2.61 & 2.07 & 1.89 & 1.85 & 1.85 & 5.17 & 4.12 & 1.58 & \textBF{1.28} \\ 
  DEN & 3.81 & 3.69 & 4.62 & 4.27 & 4.68 & 4.54 & 8.46 & 7.47 & 3.41 & \textBF{2.60} \\ 
  FIN & 5.02 & 4.93 & 6.17 & 5.37 & 5.70 & 6.19 & 8.73 & 7.84 & 3.78 & \textBF{3.42} \\ 
  FRA & 2.07 & 1.85 & 1.99 & 1.93 & 1.86 & 1.87 & 4.90 & 3.84 & 2.02 & \textBF{1.75} \\ 
  ITA & 2.33 & \textBF{1.89} & 3.21 & 2.04 & 1.99 & 2.24 & 4.38 & 3.95 & 2.10 & 2.14 \\ 
  JPN & 3.22 & 2.13 & 7.24 & 2.74 & 1.82 & 1.70 & 2.96 & 3.02 & 1.56 & \textBF{1.40} \\ 
  NET & 3.20 & 3.08 & 2.76 & 2.43 & 2.71 & 2.52 & 7.58 & 5.99 & 2.49 & \textBF{2.22} \\ 
  NZ & 5.87 & 5.75 & 4.86 & 4.68 & 4.90 & 5.63 & 9.18 & 8.19 & 5.01 & \textBF{4.63} \\ 
  NOR & 4.36 & 4.29 & 3.38 & 3.20 & 3.48 & 3.23 & 6.92 & 6.19 & 3.23 & \textBF{2.94} \\ 
  PRT & 5.34 & 4.33 & 5.54 & 4.21 & 4.58 & 4.20 & 5.57 & 4.69 & 4.24 & \textBF{3.92} \\ 
  SPA & 4.00 & 3.32 & 3.85 & 2.59 & 2.91 & \textBF{2.11} & 4.24 & 3.39 & 2.62 & 2.69 \\ 
  SWE & 3.00 & 2.93 & 2.83 & 2.76 & 2.94 & 3.18 & 8.19 & 6.44 & 2.46 & \textBF{2.14} \\ 
  SWI & 3.09 & 3.02 & 4.30 & 4.14 & 4.34 & 4.57 & 7.12 & 5.11 & \textBF{2.62} & 2.64 \\ 
  SCO & 4.79 & 4.62 & 4.01 & 3.80 & 4.12 & 4.45 & 7.67 & 7.51 & 4.12 & \textBF{3.40} \\ 
  EW & 2.52 & 2.35 & 1.96 & 1.84 & 1.86 & \textBF{1.79} & 6.49 & 5.85 & 2.14 & 1.80 \\ 
  IRE & 6.57 & 5.74 & 6.54 & \textBF{4.93} & 5.32 & 5.28 & 6.54 & 6.42 & 5.72 & 5.55 \\ 
  USA & 3.18 & 2.92 & 2.03 & 1.78 & 1.91 & 1.27 & 5.89 & 5.51 & 2.67 & \textBF{1.19} \\
\midrule 
  Mean & 3.81 & 3.48 & 3.86 & 3.18 & 3.31 & 3.29 & 6.46 & 5.55 & 3.00 & \textBF{2.68} \\ 
\\
  \underline{Male} & \\
  AUS & 5.90 & 5.53 & 4.75 & 4.44 & \textBF{4.35} & 4.57 & 11.27 & 7.63 & 4.60 & 4.57 \\ 
  BEL & 8.79 & 7.71 & 7.73 & 8.24 & 6.87 & 7.81 & 9.87 & 8.85 & 5.99 & \textBF{5.49} \\ 
  CAN & 5.31 & 4.78 & 3.86 & 3.39 & 3.62 & 3.31 & 8.68 & 6.42 & 3.55 & \textBF{2.65} \\ 
  DEN & 8.23 & 7.04 & 7.48 & 8.33 & 7.08 & 7.97 & 11.80 & 9.96 & 6.37 & \textBF{5.49} \\ 
  FIN & 10.36 & 9.95 & 9.13 & 8.63 & 9.65 & 9.15 & 13.67 & 12.30 & \textBF{5.20}  & 7.39 \\ 
  FRA & 3.04 & \textBF{2.66} & 2.91 & 2.85 & 2.99 & 3.05 & 10.62 & 8.20 & 3.32 & 2.73 \\ 
  ITA & 4.75 & 3.45 & 4.16 & 3.44 & 3.05 & 3.33 & 6.68 & 6.35 & 3.15 & \textBF{3.04} \\ 
  JPN & 3.32 & 2.52 & 4.97 & 3.62 & 3.31 & 3.85 & 4.72 & 4.76 & 2.38 & \textBF{2.22} \\ 
  NET & 7.46 & 6.26 & 5.85 & 5.12 & 5.25 & 4.35 & 8.30 & 6.69 & 5.20  & \textBF{3.79} \\ 
  NZ & 10.51 & 9.45 & 9.53 & 9.11 & 8.47 & 9.39 & 14.37 & 12.24 & 8.62 & \textBF{7.90} \\ 
  NOR & 8.36 & 7.57 & 6.95 & \textBF{5.82} & 5.92 & 6.16 & 10.21 & 9.11 & 6.39 & 5.88 \\ 
  PRT & 7.45 & 6.40 & 7.32 & 6.97 & 7.56 & 7.54 & 9.24 & 9.16 & 5.87 & \textBF{5.76} \\ 
  SPA & 5.35 & 4.32 & 4.54 & 3.80 & 4.40 & 3.45 & 7.62 & 7.21 & 3.48 & \textBF{3.37} \\ 
  SWE & 5.36 & 5.05 & 5.21 & 4.44 & 4.83 & 4.62 & 12.01 & 9.02 & 4.11 & \textBF{3.65} \\ 
  SWI & 6.23 & 6.00 & 6.76 & 6.63 & 7.01 & 7.00 & 11.77 & 8.96 & 5.15 & \textBF{4.68} \\ 
  SCO & 8.06 & 7.35 & 7.96 & 7.37 & 8.05 & 8.14 & 11.02 & 10.03 & 6.73 & \textBF{5.33} \\ 
  EW & 4.29 & 3.71 & 4.03 & 3.26 & 3.63 & 3.19 & 9.37 & 6.86 & 2.81 & \textBF{2.79} \\ 
  IRE & 15.37 & 12.27 & 11.78 & \textBF{10.10} & 10.43 & 10.13 & 13.84 & 13.32 & 11.78 & 11.12 \\ 
  USA & 5.26 & 4.75 & 3.25 & 2.98 & 3.13 & 1.91 & 11.19 & 8.33 & 4.12 & \textBF{1.85} \\ 
  \midrule
  Mean & 7.02 & 6.15 & 6.22 & 5.71 & 5.77 & 5.73 & 10.33 & 8.71 & 5.20 & \textBF{4.72} \\
\bottomrule
\end{longtable}
\end{center}

    
For the long-term forecast horizon (i.e., the 30-step-ahead forecast horizon), we also compute the mean of the MAPEs and RMSPEs to evaluate the point forecast accuracy. From Table~\ref{tab:11}, it is unclear from the comparison of the point forecast errors if there is an advantage of modeling and forecasting the truncated series for the female mortality. In contrast, there is an advantage of modeling and forecasting the male mortality and forecasting the whole series and then truncating the mortality forecasts. By comparing the mean errors of the 19 countries, Table~\ref{tab:11} shows that the most accurate forecasting method is the Lee-Carter model with Poisson errors for providing best estimates of the female mortality forecasts. The most accurate forecasting method is the APC model for providing the best estimates of the male mortality forecasts. The Lee-Carter model with Poisson errors produces smaller MAPEs and RMSPEs than the Lee-Carter model with Gaussian errors.
    
\begin{center}
\tabcolsep 0.05in
\begin{longtable}{@{}llrrrrrrrrrrrr@{}}
\caption{For the 30-step-ahead forecast horizon $h=30$, we compute the MAPE and RMSPE for each country and each model. The most accurate model for each country is highlighted in bold. For each model, we consider modeling the data with either a partial age range (termed as Partial) or a full age range (termed as Full).} \label{tab:11} \\
\toprule
Error & Country & \multicolumn{2}{c}{LC (Poisson)} & \multicolumn{2}{c}{LC (Gaussian)}  & \multicolumn{2}{c}{LC$_2$ (Gaussian)} & \multicolumn{2}{c}{APC} & \multicolumn{2}{c}{Plat} \\ 
& & Full & Partial  & Full & Partial   & Full & Partial   & Full & Partial   & Full & Partial   \\
\midrule
\endfirsthead

\toprule
Error & Country & \multicolumn{2}{c}{LC (Poisson)} & \multicolumn{2}{c}{LC (Gaussian)}  & \multicolumn{2}{c}{LC$_2$ (Gaussian)} & \multicolumn{2}{c}{APC} & \multicolumn{2}{c}{Plat} \\ 
& & Full & Partial  & Full & Partial   & Full & Partial   & Full & Partial   & Full & Partial \\
 \midrule
\endhead

\hline \multicolumn{12}{r}{{Continued on next page}} \\ 
\endfoot
\endlastfoot
MAPE & \underline{Female} & \\
& AUS & 18.78 & 20.05 & 19.09 & 20.46 & 18.62 & 21.58 & \textBF{15.20} & 26.11 & 20.53 & 26.99 \\ 
&   BEL & \textBF{15.19} & 15.50 & 17.02 & 15.99 & 20.22 & 16.61 & 15.77 & 21.65 & 26.27 & 29.78 \\ 
&   CAN & 13.21 & 12.72 & 12.56 & 9.93 & \textBF{9.07} & 10.68 & 17.96 & 16.29 & 22.76 & 10.32 \\ 
&   DEN & 11.91 & 12.36 & 11.44 & \textBF{10.71} & 14.64 & 11.41 & 23.59 & 22.03 & 34.15 & 19.92 \\ 
&   FIN & \textBF{16.35} & 18.28 & 16.79 & 18.37 & 21.59 & 20.92 & 23.38 & 27.27 & 26.10 & 29.23 \\ 
&   FRA & 16.81 & 17.35 & 16.74 & 17.30 & 20.00 & 18.23 & \textBF{13.80} & 18.51 & 19.02 & 30.02 \\ 
&   ITA & 18.55 & 21.17 & 26.00 & 24.22 & 24.16 & 22.90 & \textBF{15.06} & 32.54 & 23.41 & 48.43 \\ 
&   JPN & 22.83 & 23.89 & 40.71 & 27.40 & 24.65 & 23.03 & 52.48 & \textBF{19.82} & 52.96 & 30.09 \\ 
&   NET & 15.04 & 14.42 & 20.25 & 14.41 & \textBF{11.70} & 12.82 & 14.91 & 16.98 & 31.19 & 15.17 \\ 
&   NZ & 34.61 & 35.91 & 38.23 & 40.11 & 38.56 & 36.21 & \textBF{28.84} & 41.59 & 35.94 & 44.18 \\ 
&   NOR & 10.17 & \textBF{9.52} & 12.32 & 11.39 & 10.44 & 10.44 & 14.22 & 19.68 & 35.73 & 12.53 \\ 
&   PRT & 41.89 & 34.61 & 53.01 & 40.15 & 52.41 & 45.16 & \textBF{17.40} & 46.54 & 47.46 & 54.60 \\ 
&   SPA & 25.09 & \textBF{24.33} & 52.66 & 27.03 & 27.00 & 28.85 & 35.96 & 27.82 & 35.92 & 40.34 \\ 
&   SWE & 14.54 & 13.73 & 15.85 & 12.50 & 13.50 & 12.55 & 16.02 & 16.26 & 24.55 & \textBF{12.05} \\ 
&   SWI & 11.27 & 10.93 & 10.99 & \textBF{10.83} & 12.25 & 12.30 & 17.56 & 19.76 & 37.25 & 18.94 \\ 
&   SCO & 16.91 & 18.29 & 17.55 & 19.72 & 19.75 & 20.56 & \textBF{10.77} & 25.86 & 31.53 & 20.71 \\ 
&   EW & 17.24 & 19.41 & 55.25 & 19.25 & 17.46 & 20.11 & \textBF{9.57} & 34.12 & 25.03 & 34.45 \\ 
&   IRE & 38.93 & 38.09 & 68.20 & 43.59 & 45.58 & 41.77 & \textBF{19.47} & 44.09 & 38.19 & 41.49 \\ 
&   USA & 8.90 & 7.51 & 11.13 & 5.98 & 7.75 & \textBF{5.55} & 16.65 & 13.24 & 12.56 & 7.83 \\ 
\cmidrule{2-12}
&   Mean & 19.38 & \textBF{19.37} & 27.15 & 20.49 & 21.54 & 20.62 & 19.93 & 25.80 & 30.56 & 27.74 \\ 
\\
&  \underline{Male} & \\
& AUS & 50.13 & 52.47 & 56.68 & 55.41 & 49.33 & 55.95 & \textBF{38.21} & 52.54 & 53.81 & 72.66 \\ 
&   BEL & 58.34 & 54.88 & 65.69 & 92.47 & 58.83 & 85.97 & \textBF{31.63} & 46.42 & 87.24 & 46.11 \\ 
&   CAN & 45.46 & 42.26 & 47.85 & 44.46 & 47.80 & 44.78 & \textBF{33.70} & 43.26 & 79.94 & 49.49 \\ 
&   DEN & 63.45 & 63.01 & 63.36 & 67.67 & 66.87 & 67.44 & 66.01 & 57.71 & 73.57 & \textBF{56.30} \\ 
&   FIN & 43.39 & 48.11 & 44.34 & 96.58 & 47.78 & 51.77 & \textBF{20.85} & 31.29 & 75.96 & 52.26 \\ 
&   FRA & \textBF{32.81} & 33.83 & 35.81 & 34.77 & 37.00 & 35.36 & 39.09 & 35.86 & 60.49 & 39.45 \\ 
&   ITA & 64.41 & 62.35 & 70.38 & 67.22 & 53.63 & 62.03 & \textBF{32.11} & 61.05 & 81.64 & 60.60 \\ 
&   JPN & \textBF{12.30} & 12.68 & 20.98 & 12.79 & 12.91 & 12.82 & 39.53 & 20.94 & 32.12 & 15.61 \\ 
&   NET & 79.68 & 76.14 & 78.67 & 79.29 & 65.19 & 71.08 & \textBF{45.62} & 51.89 & 79.88 & 47.61 \\ 
&   NZ & 99.32 & 82.86 & 112.42 & 117.61 & 104.32 & 114.28 & \textBF{63.10} & 76.39 & 92.05 & 94.38 \\ 
&   NOR & 87.50 & 73.97 & 88.62 & 71.43 & 65.64 & 67.35 & \textBF{45.09} & 66.85 & 70.00 & 60.19 \\ 
&   PRT & 42.79 & 38.11 & 49.01 & 83.79 & 58.24 & 49.59 & \textBF{18.01} & 35.29 & 32.26 & 47.62 \\ 
&   SPA & 25.43 & 23.13 & 58.45 & 29.85 & 27.15 & 30.84 & \textBF{15.21} & 25.02 & 26.29 & 28.47 \\ 
&   SWE & 51.89 & 47.57 & 55.01 & 57.89 & 53.05 & 51.91 & 45.46 & 43.28 & 76.65 & \textBF{42.27} \\ 
&   SWI & 36.00 & 37.46 & 41.73 & 42.27 & 39.01 & 45.48 & \textBF{30.64} & 46.42 & 50.82 & 55.93 \\ 
&   SCO & 52.66 & 55.63 & 55.86 & 92.86 & 61.36 & 86.41 & \textBF{38.51} & 57.02 & 65.63 & 54.61 \\ 
&   EW & 52.44 & 55.96 & 52.81 & 60.03 & 50.06 & 55.52 & \textBF{35.42} & 48.16 & 47.26 & 54.41 \\ 
&   IRE & 110.20 & 91.02 & 112.71 & 111.84 & 111.82 & 104.52 & \textBF{43.10} & 97.26 & 63.07 & 93.34 \\ 
&   USA & 24.07 & 24.99 & 26.07 & 26.81 & 26.12 & 28.13 & \textBF{21.56} & 23.20 & 70.21 & 31.07 \\ 
  \cmidrule{2-12}
& Mean & 54.33 & 51.39 & 59.81 & 65.53 & 54.53 & 59.01 & \textBF{36.99} & 48.41 & 64.15 & 52.76 \\ 
\cmidrule{2-12}
RMSPE & \underline{Female} & \\
 & AUS & 21.74 & 23.11 & 22.17 & 23.51 & 21.82 & 24.60 & \textBF{19.09} & 28.34 & 24.41 & 30.76 \\ 
 &   BEL & \textBF{18.25} & 18.92 & 20.82 & 19.44 & 24.70 & 20.38 & 18.67 & 24.89 & 31.54 & 35.64 \\ 
 &   CAN & 15.37 & 14.95 & 14.53 & 12.18 & \textBF{11.31} & 13.01 & 22.76 & 21.38 & 28.37 & 14.13 \\ 
 &   DEN & 14.08 & 14.44 & \textBF{13.33} & 13.46 & 16.62 & 13.50 & 25.58 & 25.11 & 40.84 & 23.08 \\ 
 &   FIN & \textBF{19.82} & 22.37 & 20.14 & 22.67 & 26.18 & 25.80 & 27.37 & 30.78 & 33.02 & 35.66 \\ 
 &   FRA & 19.01 & 19.60 & 19.27 & 20.17 & 23.47 & 21.18 & \textBF{17.97} & 21.60 & 24.55 & 34.67 \\ 
 &   ITA & 21.18 & 23.45 & 28.25 & 26.36 & 26.44 & 25.17 & \textBF{18.28} & 36.15 & 28.19 & 57.57 \\ 
 &   JPN & 25.73 & 26.91 & 47.72 & 30.06 & 27.27 & \textBF{25.71} & 53.75 & 27.37 & 55.50 & 33.61 \\ 
 &   NET & 16.33 & 15.77 & 21.13 & 15.67 & \textBF{13.93} & 14.43 & 20.15 & 21.32 & 37.43 & 17.97 \\ 
 &   NZ & 44.37 & 45.84 & 46.55 & 49.71 & 46.68 & 45.70 & \textBF{33.10} & 48.79 & 42.29 & 50.98 \\ 
 &   NOR & 12.65 & \textBF{12.45} & 14.53 & 13.90 & 12.80 & 13.13 & 17.94 & 22.61 & 41.08 & 14.97 \\ 
 &   PRT & 48.09 & 40.40 & 61.17 & 46.28 & 60.66 & 52.06 & \textBF{20.96} & 54.09 & 52.80 & 62.86 \\ 
 &   SPA & 30.73 & \textBF{29.51} & 59.17 & 33.96 & 33.89 & 36.16 & 38.24 & 30.74 & 42.10 & 45.40 \\ 
 &   SWE & 16.23 & 15.50 & 17.54 & 14.61 & 17.18 & 14.73 & 22.05 & 22.21 & 31.04 & \textBF{13.67} \\ 
  &   SWI & 13.97 & 13.52 & 13.60 & \textBF{13.42} & 15.17 & 17.27 & 23.51 & 24.38 & 42.62 & 21.26 \\ 
   & SCO & 22.00 & 24.51 & 20.56 & 25.95 & 22.19 & 26.98 & \textBF{14.05} & 29.29 & 35.82 & 24.81 \\ 
 &   EW & 21.84 & 24.08 & 61.19 & 23.89 & 21.62 & 24.44 & \textBF{11.78} & 38.79 & 29.59 & 39.16 \\ 
 &   IRE & 44.06 & 43.59 & 73.27 & 48.92 & 50.75 & 47.72 & \textBF{24.33} & 57.14 & 42.55 & 47.34 \\ 
 &   USA & 10.10 & 8.70 & 12.13 & 7.02 & 9.11 & \textBF{6.77} & 18.45 & 16.32 & 15.29 & 8.77 \\ 
 \cmidrule{2-12}
 &   Mean & \textBF{22.92} & 23.03 & 30.90 & 24.27 & 25.36 & 24.67 & 23.58 & 30.60 & 35.74 & 32.23 \\ 
 \\
&  \underline{Male} & \\
& AUS & 58.82 & 61.48 & 66.43 & 64.80 & 57.73 & 66.12 & \textBF{45.92} & 64.35 & 64.19 & 82.67 \\ 
&   BEL & 72.35 & 63.85 & 78.04 & 103.40 & 71.52 & 95.86 & \textBF{37.45} & 54.36 & 90.56 & 53.55 \\ 
&   CAN & 57.57 & 52.42 & 61.40 & 55.54 & 60.60 & 55.81 & \textBF{40.51} & 52.40 & 84.49 & 57.41 \\ 
&   DEN & 81.11 & 80.89 & 78.71 & 75.38 & 82.53 & 82.03 & 77.00 & \textBF{66.75} & 91.52 & 68.67 \\ 
&   FIN & 51.71 & 57.52 & 52.22 & 111.49 & 55.40 & 61.78 & \textBF{23.39} & 36.30 & 78.51 & 60.90 \\ 
&   FRA & \textBF{38.15} & 38.84 & 41.53 & 40.38 & 40.58 & 40.91 & 43.68 & 40.64 & 68.48 & 44.68 \\ 
&   ITA & 80.70 & 77.32 & 87.79 & 84.15 & 67.36 & 78.15 & \textBF{36.01} & 71.62 & 85.65 & 75.05 \\ 
&   JPN & 15.46 & 16.48 & 25.77 & \textBF{15.25} & 16.66 & 15.91 & 40.62 & 26.82 & 37.06 & 18.33 \\ 
&   NET & 102.58 & 95.23 & 100.24 & 102.76 & 81.69 & 89.74 & \textBF{53.64} & 60.81 & 87.97 & 64.18 \\ 
&   NZ & 112.83 & 98.95 & 128.69 & 136.49 & 117.87 & 132.75 & \textBF{74.82} & 91.95 & 102.96 & 107.61 \\ 
&   NOR & 108.62 & 85.81 & 110.96 & 82.25 & 80.62 & 79.43 & \textBF{50.60} & 77.04 & 79.85 & 76.25 \\ 
&   PRT & 49.93 & 44.03 & 56.63 & 95.17 & 64.89 & 55.57 & \textBF{20.40} & 41.52 & 38.67 & 55.07 \\ 
&   SPA & 32.82 & 30.04 & 63.13 & 33.14 & 32.15 & 34.60 & \textBF{19.17} & 27.51 & 30.23 & 33.06 \\ 
&   SWE & 62.19 & 56.65 & 66.37 & 68.98 & 62.41 & 62.39 & 52.86 & 51.01 & 84.63 & \textBF{50.20} \\ 
&   SWI & 43.71 & 45.27 & 48.84 & 50.17 & 45.93 & 53.22 & \textBF{36.91} & 55.66 & 61.52 & 60.67 \\ 
&   SCO & 64.25 & 66.37 & 68.29 & 105.02 & 72.18 & 97.90 & \textBF{44.12} & 67.61 & 71.81 & 66.18 \\ 
&   EW & 62.45 & 65.98 & 63.72 & 71.74 & 58.80 & 65.56 & \textBF{42.07} & 58.11 & 56.89 & 63.73 \\ 
&   IRE & 122.60 & 104.66 & 125.27 & 128.85 & 124.03 & 118.34 & \textBF{49.25} & 116.02 & 80.35 & 105.59 \\ 
&   USA & 28.44 & 29.33 & 30.26 & 31.41 & 30.41 & 32.05 & \textBF{25.85} & 26.09 & 72.70 & 34.40 \\ 
\cmidrule{2-12}
&  Mean & 65.59 & 61.64 & 71.28 & 76.65 & 64.39 & 69.37 & \textBF{42.86} & 57.19 & 72.00 & 62.01 \\ 
\bottomrule
\end{longtable}
\end{center}

For the long-term forecast horizon (i.e., the 30-step-ahead forecast horizon), we also compute the $\overline{S}_{\alpha=0.2}$ to evaluate the interval forecast accuracy. From Table~\ref{tab:33}, there is a slight advantage of directly modeling and forecasting the truncated series for the female mortality. For modeling the male mortality, there is an advantage of modeling and forecasting the whole series and then truncating the mortality forecasts. By comparing the mean errors of the 19 countries, Table~\ref{tab:33} shows that the most accurate forecasting method is the Lee-Carter model with Poisson errors for providing the interval forecasts of the female mortality rates. The most accurate forecasting method is the APC model for providing the interval forecasts of the male mortality rates. The Lee-Carter model with Poisson errors produces smaller MAPEs than the Lee-Carter model with Gaussian errors.

\begin{center}
\begin{longtable}{@{}lrrrrrrrrrrrr@{}}
\caption{For the 30-step-ahead forecast horizon $h=30$, we compute the mean interval score ($\times 100$) for each country and each model.} \label{tab:33} \\
\toprule
Country & \multicolumn{2}{c}{LC (Poisson)} & \multicolumn{2}{c}{LC (Gaussian)}  & \multicolumn{2}{c}{LC$_2$ (Gaussian)} & \multicolumn{2}{c}{APC} & \multicolumn{2}{c}{Plat} \\ 
 & Full & Partial  & Full & Partial   & Full & Partial   & Full & Partial   & Full & Partial \\
 \midrule
\endfirsthead

\toprule
Country & \multicolumn{2}{c}{LC (Poisson)} & \multicolumn{2}{c}{LC (Gaussian)}  & \multicolumn{2}{c}{LC$_2$ (Gaussian)} & \multicolumn{2}{c}{APC} & \multicolumn{2}{c}{Plat} \\ 
 & Full & Partial  & Full & Partial   & Full & Partial   & Full & Partial   & Full & Partial \\
  \midrule
\endhead

\hline \multicolumn{11}{r}{{Continued on next page}} \\ 
\endfoot
\endlastfoot
\underline{Female} & \\
AUS & 8.25 & \textBF{7.20} & 9.30 & 8.94 & 7.75 & 8.80 & 13.42 & 10.46 & 12.31 & 9.29 \\ 
  BEL & 5.99 & \textBF{4.99} & 6.85 & 6.18 & 7.83 & 10.10 & 8.24 & 11.80 & 19.53 & 12.75 \\ 
  CAN & 12.64 & 10.76 & 7.54 & 8.09 & 5.11 & 7.09 & 23.15 & 21.93 & 23.71 & \textBF{4.47} \\ 
  DEN & \textBF{4.81} & 5.23 & 5.37 & 5.07 & 6.35 & 18.66 & 9.72 & 8.85 & 28.91 & 8.85 \\ 
  FIN & 12.80 & 9.57 & \textBF{9.23} & 9.29 & 12.38 & 162.23 & 25.85 & 20.57 & 27.49 & 16.37 \\ 
  FRA & 5.39 & \textBF{4.51} & 5.49 & 7.33 & 5.45 & 5.69 & 11.37 & 11.82 & 14.18 & 9.99 \\ 
  ITA & 8.81 & \textBF{4.54} & 12.62 & 10.21 & 6.16 & 8.46 & 6.16 & 7.09 & 7.25 & 13.39 \\ 
  JPN & 6.73 & \textBF{4.30} & 12.47 & 10.43 & 5.40 & 6.06 & 27.24 & 20.42 & 39.63 & 7.77 \\ 
  NET & 9.94 & 9.09 & 10.61 & 9.66 & \textBF{6.96} & 8.11 & 22.06 & 23.59 & 42.29 & 8.94 \\ 
  NZ & 8.75 & 8.81 & 9.61 & 10.39 & 16.15 & 82.21 & \textBF{7.78} & 8.99 & 9.13 & 13.52 \\ 
  NOR & 6.81 & 7.82 & 5.39 & 6.41 & \textBF{5.21} & 7.68 & 14.47 & 13.55 & 44.89 & 7.80 \\ 
  PRT & 20.63 & 10.59 & 26.58 & 13.21 & 23.64 & 18.53 & 8.51 & \textBF{8.49} & 53.85 & 22.47 \\ 
  SPA & 16.70 & 13.35 & 18.08 & 16.12 & 12.87 & 9.29 & 15.95 & 12.68 & 32.88 & \textBF{8.74} \\ 
  SWE & 9.66 & 8.88 & 9.47 & 7.85 & \textBF{6.16} & 13.71 & 25.97 & 26.32 & 35.74 & 6.55 \\ 
  SWI & 3.67 & \textBF{3.39} & 5.30 & 5.08 & 6.58 & 42.74 & 22.61 & 18.84 & 31.94 & 14.23 \\ 
  SCO & 8.36 & \textBF{6.08} & 9.65 & 7.65 & 10.37 & 30.24 & 7.13 & 6.64 & 26.02 & 11.29 \\ 
  EW & \textBF{3.84} & 4.04 & 11.42 & 4.45 & 4.81 & 6.20 & 5.09 & 7.44 & 13.97 & 8.55 \\ 
  IRE & 17.25 & 12.39 & 20.57 & 16.64 & 18.89 & 21.43 & \textBF{8.06} & 9.81 & 15.26 & 20.33 \\ 
  USA & 4.11 & 2.50 & 3.11 & \textBF{1.80} & 2.52 & 2.70 & 15.66 & 13.93 & 4.47 & 4.05 \\ 
\midrule
  Mean & 9.22 & \textBF{7.27} & 10.46 & 8.67 & 8.98 & 24.73 & 14.65 & 13.85 & 25.45 & 11.02 \\ 
\\
  \underline{Male} & \\
  AUS & 15.97 & 12.78 & 18.91 & 15.69 & 13.62 & 23.85 & \textBF{10.15} & 10.18 & 16.51 & 21.01 \\ 
  BEL & 22.28 & 18.18 & 29.43 & 47.28 & 197.68 & 274.40 & \textBF{9.25} & 12.30 & 39.35 & 17.42 \\ 
  CAN & 12.99 & \textBF{11.43} & 12.53 & 11.80 & 12.94 & 13.63 & 13.50 & 13.63 & 35.64 & 14.97 \\ 
  DEN & 19.74 & 19.28 & 21.72 & 30.66 & 441.92 & 463.81 & 22.09 & 16.40 & 23.17 & \textBF{15.61} \\ 
  FIN & 16.45 & 16.98 & 15.54 & 33.36 & 26.11 & 43.94 & \textBF{8.54} & 8.80 & 34.82 & 23.53 \\ 
  FRA & 10.58 & \textBF{8.20} & 12.85 & 12.74 & 14.71 & 14.93 & 10.69 & 9.38 & 16.70 & 14.70 \\ 
  ITA & 17.34 & 16.24 & 19.64 & 17.47 & 12.56 & 17.47 & \textBF{7.89} & 15.92 & 27.76 & 17.58 \\ 
  JPN & \textBF{3.58} & 3.69 & 8.70 & 4.73 & 7.27 & 23.37 & 30.41 & 25.32 & 38.02 & 10.07 \\ 
  NET & 23.74 & 21.63 & 24.04 & 20.78 & 18.91 & 23.43 & 14.96 & \textBF{10.42} & 28.78 & 15.71 \\ 
  NZ & 45.00 & 26.13 & 54.39 & 41.50 & 52.59 & 95.29 & 18.23 & \textBF{16.61} & 28.10 & 42.92 \\ 
  NOR & 26.45 & 25.79 & 25.86 & 27.40 & 16.89 & 23.49 & 14.77 & \textBF{13.63} & 22.20 & 16.68 \\ 
  PRT & 19.87 & 13.68 & 24.01 & 37.26 & 30.92 & 118.32 & \textBF{8.50} & 12.32 & 12.26 & 22.79 \\ 
  SPA & 27.44 & 14.80 & 35.05 & 19.88 & 15.81 & 14.26 & \textBF{8.92} & 10.80 & 11.09 & 10.42 \\ 
  SWE & 17.71 & 13.46 & 18.53 & 17.60 & 19.87 & 20.01 & 14.96 & 11.34 & 25.19 & \textBF{10.65} \\ 
  SWI & 9.37 & 9.05 & 14.48 & 13.32 & 28.57 & 28.63 & \textBF{9.04} & 11.19 & 14.00 & 22.49 \\ 
  SCO & 20.85 & 19.01 & 21.61 & 43.03 & 84.51 & 96.68 & \textBF{10.22} & 12.77 & 25.07 & 18.27 \\ 
  EW & 15.24 & 14.20 & 16.68 & 21.06 & 17.35 & 17.39 & \textBF{7.48} & 11.17 & 14.08 & 17.50 \\ 
  IRE & 70.26 & 37.26 & 73.40 & 48.79 & 69.77 & 109.01 & \textBF{12.71} & 25.21 & 28.22 & 36.51 \\ 
  USA & 10.95 & 9.65 & 10.60 & \textBF{9.26} & 10.47 & 10.34 & 12.59 & 11.09 & 42.64 & 11.69 \\ 
\midrule
   Mean & 21.36 & 16.39 & 24.10 & 24.93 & 57.50 & 75.38 & \textBF{12.89} & 13.60 & 25.45 & 18.97 \\ 
\bottomrule
\end{longtable}
\end{center}


\section{Conclusion}\label{sec:6}

We consider forecasting retiree mortality using two modeling strategies. On the one hand, we can first truncate the available data to retiree ages and then produce mortality forecasts. On the other hand, we can first use the available data to produce forecasts and then truncate the mortality forecasts to retiree ages. Using the empirical data from \cite{HMD19}, we investigate the short-term and long-term point and interval forecast accuracies. Between the two model strategies, we recommend the first strategy by truncating all available data to retiree ages and then produce short-term mortality forecasts. Our recommendations could be useful to actuaries for choosing a better modeling strategy and more accurately pricing a range of annuity products.

For the long-term mortality forecasts, we recommend the first strategy for modeling female mortality but the second strategy for modeling male mortality. It is difficult to recommend a strategy when the model and its parameters may not be optimal for the long-term forecasts. This is a disadvantage of using methods based on time series extrapolation for long term forecasting. Instead, an expectation approach, in which experts set a future target, could be considered, noting that this method has also had limited success in the past \citep{BT08}.

There are several ways in which the present study can be further extended, and we briefly mention two: 
\begin{inparaenum}
\item[1)] We could consider some machine learning methods to model mortality forecasts \citep[see, e.g.,][]{RW20, PRS+20}.
\item[2)] The results depend on the age range considered as well as the selected countries. 
\end{inparaenum}
The R code for reproducing the results can be provided upon request from the corresponding author.

\section*{Acknowledgments}

The authors are grateful to the comments and suggestions received from three reviewers, and the scientific committee of the Living to 100 symposium in 2020.

\newpage
\bibliographystyle{agsm}
\bibliography{partial_or_full.bib}

\end{document}